# Fe-rich border of miscibility gap and activation energy of phase decomposition in a Fe-Cr alloy

S. M. Dubiel[*] and J. Żukrowski

*AGH University of Science and Technology, Faculty of Physics and Applied Computer Science, al. A. Mickiewicza 30, 30-059 Kraków, Poland*

**Abstract**

Concentration of Cr in the Fe-rich $\alpha$-phase, $x$, resulted from a phase decomposition caused by an isothermal annealing at T = 415 and 450 $^o$C of a non-irradiated (NR) Fe-Cr14 EFDA sample and that of a He-ions irradiated (IR) one annealed at 415 $^o$C was determined with Mössbauer spectroscopy. The $x$-value in the latter was by ~3 at% higher than the one in the NR-counterpart. The activation energy for the phase decomposition in the NR-sample was 122 kJ/mol. In the IR-sample its value was by 12 kJ/mol lower. Avrami exponents for the NR-samples were close to 0.5, and that for the IR-sample had a value of about 1.

* Corresponding author: **Stanislaw.Dubiel@fis.agh.edu.pl**

## 1. Introduction

Phase decomposition into Fe-rich (α) and Cr-rich (α') phases is the main reason for the so-called *475°C embrittlement* that may occur in technologically important structural materials made on the base of Fe-Cr alloys such as ferritic stainless steels, if subjected to temperatures between ~300 and ~600 °C [1]. This frequently occurs in practice, as different devices fabricated from such materials (e. g. heat exchangers, gas turbines, etc.) work at service at elevated temperatures. The α' precipitation leads to a progressive deterioration of materials' useful properties like hardening and reduction of their toughness. In the phase-diagram of the Fe-Cr system, the region where the decomposition may occur is known as *the miscibility gap*. Knowledge of its borders, especially of its Fe-rich line, is of a vital importance for predicting and understanding of the steels behaviour at elevated temperatures. Theoretically calculated phase diagrams of the Fe-Cr alloy system are at variance with each other, in particular, as far as the borders of the miscibility gap are concerned [2-5]. Their verification above ~430 °C is hardly possible due to a wide scatter of the experimental data in that range of temperature, and impossible below ~430 °C due to a lack of such data available in the literature [4]. This situation justifies and prompts further relevant measurements. One of the techniques that can supply precise information on the issue is the Mössbauer spectroscopy (MS) as shown elsewhere [6]. Here, results concerning the Fe-rich line of the miscibility gap obtained with MS for non-irradiated and He-ions irradiated samples of Fe-Cr14 EFDA model alloy as well as the kinetics of the phase decomposition are reported. The effect of the irradiation is of interest, too, as the Fe-Cr alloys constitute the main ingredient of ferritic steels (FS) which are regarded as good construction materials for a new generation (IV) of nuclear power facilities [7]. At service they often are subject to irradiation which causes a radiation damage that can seriously deteriorate their mechanical properties. On the lattice scale, the radiation causes lattice defects, and, consequently, a redistribution of Fe/Cr atoms that can lead to a short-range order or the phase decomposition.

Values of the activation energy for the phase decomposition in the non-irradiated and irradiated samples as determined from the measured quantities as well as those of the Avrami exponent are also presented.

## 2. Samples and spectra measurements

As samples, a model EFDA alloy of Fe-Cr14 (chromium content 15.1 at%) was used in form of ~20 x 20 mm$^2$ plates with a thickness of ~0.2 mm. The plates were obtained by cold-rolling of slabs cut from original ingots. One of the plates was irradiated at the Helmholtz-Zentrum Dresden-Rossendorf, Germany, with 25 keV He-ions to the final dose of $1.2 \cdot 10^{17}$ He$^+$/cm$^2$ (7.5 dpa). The temperature of the sample during the irradiation was less than 100 °C., and the implantation depth ~225 nm. The phase decomposition process was caused by isothermal annealing of the samples in vacuum at two temperatures: 415 °C (non-irradiated and irradiated samples) and 450 °C (non-irradiated sample). $^{57}$Fe Mössbauer spectra, of which some examples are shown in Figure 1 (left panel), were measured *ex-situ* at room temperature recording 7.3 keV conversion electrons (CEMS mode) on the irradiated side of the samples. These electrons give information on ~300 nm thick presurface layer.

## 3. Spectra analysis and results

The measured spectra were analysed in terms of the hyperfine magnetic field distribution method described in detail elsewhere [8]. A linear relationship between the hyperfine field and the isomer shift was assumed, as found experimentally to hold [9]. The fitting procedure yielded hyperfine magnetic field distribution curves, examples of which can be seen in Figure 1 (right panel). Their integration gave an average hyperfine field, *<B>*, the quantity pertinent to determination of the alloy composition as shown elsewhere [6], was calculated. Examples of the dependence of *<B>* on time of annealing, *t*, are presented in Figs. 2 and 3. They show a characteristic saturation-like behaviour, and could be well fitted with the following Johnson-Mehl-Avrami-Kolmogorov equation:

$$<B> = <B_o> + b[1 - \exp(-kt^n)] \qquad (1)$$

Where $<B_o>$ is the value of the average hyperfine field for the non-annealed sample, $k$ is the rate constant, $n$ is the Avrami exponent, and $b$ is a free parameter.

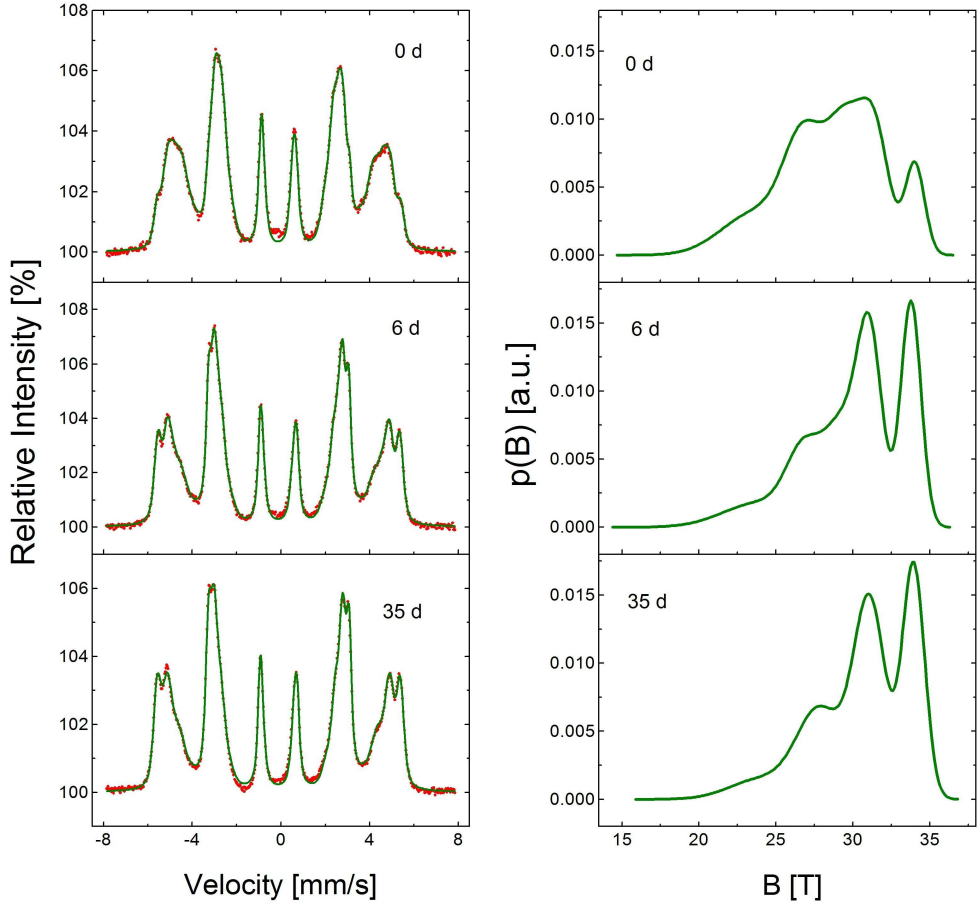

**Figure 1** (left panel) CEMS $^{57}$Fe spectra recorded at room temperature on the non-irradiated Fe-Cr14 sample without annealing (top) and annealed at 415 $^o$C for 6 days (middle) and 35 days (bottom). The solid lines are for the best fit to the data, and (right panel) corresponding distribution curves of the hyperfine magnetic field derived form the spectra.

The value of $<B>$ in saturation, $<B_S>$, can be used for the unique determination of the composition of the Fe-rich phase based on the monotonous and smooth relationship between $<B>$ and the chromium content, $x$, as reported elsewhere [6]. The value of $t$ at which the saturation state has been achieved, $t_S$, gives the information on the time necessary for completing the process of decomposition. Table I displays the two quantities for the studied samples together with the concentration of chromium in the α-phase, $x$, derived from the $<B>=f(x)$ dependence as described in detail elsewhere [6]. It is clear that (i) the actual value of $x$ for the non-irradiated samples meaningfully depends on the annealing temperature, and (2) the performed irradiation significantly affected the process of the phase decomposition as it drastically accelerated it, and also changed the concentration of chromium in the Fe-rich phase. Also the rate constant, $k$, and the Avrami exponent, $n$, as obtained by fitting equ. (1) to $<B(t)>$ data are included in the Table. Noteworthy, the decomposition in the irradiated sample is by a factor of twenty faster than that in the non-irradiated counterpart which can be understood in terms of radiation demage caused by the irradiation.

**Table I** Annealing temperature, $T_a$, saturation hyperfine field, $<B_S>$, annealing time to achieve the saturation state, $t_S$, chromium content in the Fe-rich phase, $x$, the rate constant, $k$, and the Avrami exponent, $n$, for non-irradiated (NR) and He-ions irradiated (IR) samples of Fe-Cr14.

| Sample | $T_a$ [°C] | $<B_S>$ [T] | $t_S$ [h] | x [at% Cr] | k [h$^{-1}$] | n |
|---|---|---|---|---|---|---|
| NR | 415 | 30.5(3) | ~500 | 10.5(5) | 0.5(1) | 0.6(1) |
| IR | 415 | 29.6(1) | ~25 | 13.5(2) | 3.9(9) | 1.0(3) |
| NR | 450 | 29.8(2) | ~150 | 13.3(4) | 1.4(3) | 0.44(7) |

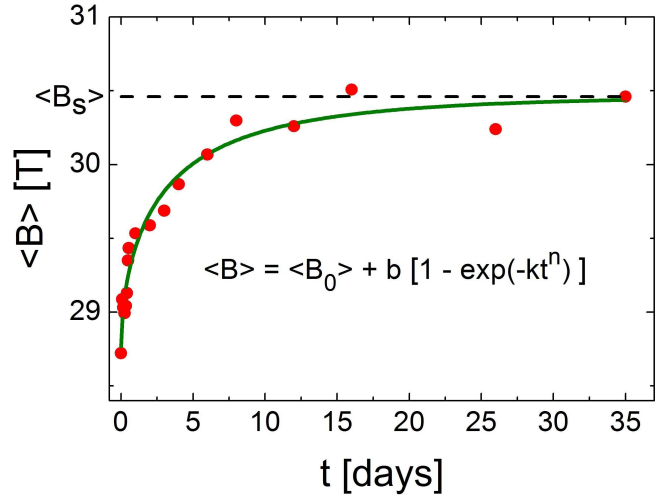

**Figure 2** Dependence of the average hyperfine field, $<B>$, on the annealing time, $t$, for the non-irradiated sample of Fe-Cr14 annealed at 415 °C. The best fit curve to the data in terms of equation (1) is indicated by a solid line. The dashed line is for $<B_S>$.

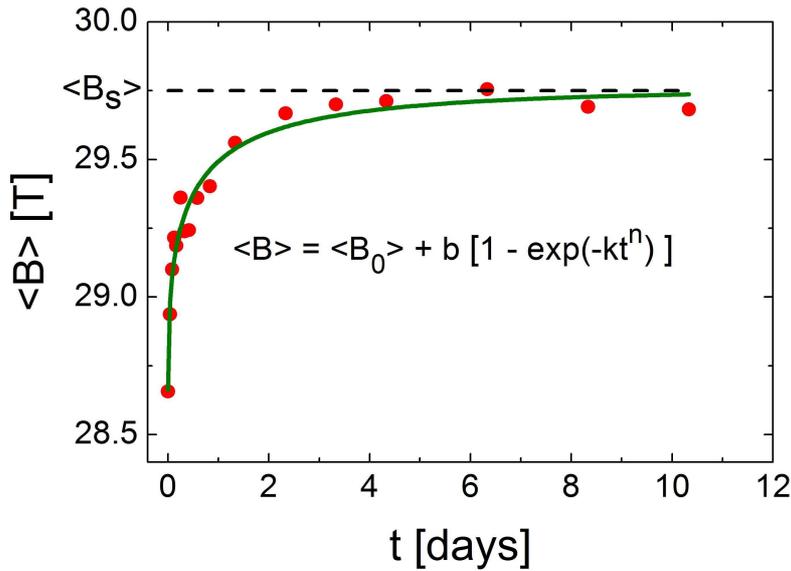

**Figure 3** Dependence of the average hyperfine field, $<B>$, on the annealing time, $t$, for the non-irradiated sample of Fe-Cr14 annealed at 450 °C. The best fit curve to the data in terms of equation (1) is indicated by a solid line. The dashed line is for $<B_S>$.

The knowledge of the rate constants, $k$, for two different temperatures, $T_1 = 415\ ^oC$ and $T_2 = 450\ ^oC$ combined with the Arrhenius law, enabled determination of the activation energy, $E$, for the phase decomposition in the temperature interval of 415–450 $^oC$, using the following equation:

$$E = \frac{T_1 T_2}{T_2 - T_1} k_B \ln\left(\frac{k_2}{k_1}\right) \qquad (2)$$

Where $k_B$ is the Boltzmann constant. The equation is valid assuming the pre-exponent $k_o$ in the Arrhenius law is temperature independent.

For the non-irradiated sample $E$ was estimated as equal to 122(6) kJ/mol (1.26(6) eV) which is significantly less the value of the activation energy for the σ-phase formation in the Fe-Cr system viz. 196(2) kJ/mol as found elsewhere [10]. The knowledge of $k$-values for the non-irradiated and irradiated samples annealed at 415$^o$C made it possible to find that the activation energy for the decomposition in the irradiated sample was by 12 kJ/mol (0.12 eV) lower (an assumption was made that the pre-exponent factor in the Arrhenius law was the same for the non- irradiated and the irradiated samples). This is in accord with the observation that the process of the phase decomposition in that sample was much faster than the one in the non-irradiated sample. The value of $n$ is usually used to analyze a mechanism responsible for the transformation. Its values obtained in this study for the non-irradiated samples, close to 0.5 for both temperatures, indicate that the process responsible for the phase decomposition in these samples could be a diffusion-controlled thickening of plates after their edges have impinged [11]. For the irradiated sample $n$ is close to 1, hence in this case the phase decomposition could have been governed by different transformation conditions, namely a growth of isolated platelets or needles of finite size of the Cr-enriched phase. Whatever the conditions were liable for the phase decomposition in the studied samples, the results obtained in this study gave a clear evidence that the He-ions irradiation significantly affected the concentration of the Fe-rich phase, drastically accelerated the kinetics of the decomposition as well as it changed the mechanism of the latter.

**4. Conclusions**

The results reported in this paper can be concluded as follows:
(1) Concentration of chromium in the Fe-rich phase (α) of the Fe-Cr14 sample annealed at 415 $^oC$ is equal to 10.5 at% while that found in α after annealing at 450 $^oC$ is equal to 13.3 at%.
(2) For the He[+]-ions irradiated sample the Fe-rich border at 415$^o$C is shifted by ~3 at% towards a higher Cr content i.e. the miscibilityt gap is narrower.
(3) The activation energy for the phase decomposition in the non-irradiated sample was estimated as equal to 122 kJ/mol (1.26 eV), while its value for the irradiated sample was by 12 kJ/mol lower.
(4) Values of Avrami exponents obtained for the non-irradiated samples are close to 0.5, hence they indicate that the thickening of platelets after their edges have impinged might be responsible for the decomposition in these samples.
(5) Avrami exponent found for the irradiated sample has a value of 1, what means that in the irradiated samples the phase decomposition is giverned by a different mechanism.

**Acknowledgements**

This work, supported by the European Communities under the contract of Association between EURATOM and IPPLM, was carried out within the framework of the European Fusion Development Agreement, and it was also supported by the Ministry of Science and Higher Education, Warszawa. H. Reuther, Helmholtz-Zentrum Dresden-Rossendorf, Germany, is acknowledged for performing the irradiation.